\documentclass[prb,twocolumn,aps,showpacs]{revtex4}
\usepackage{amssymb}
\usepackage{amsmath}
\usepackage{graphicx}
\usepackage{dcolumn}
\usepackage{bm}

\begin{document}


\title{ELECTRICAL EXPRESSION OF SPIN ACCUMULATION IN FERROMAGNET/SEMICONDUCTOR STRUCTURES}

\author{{\L}ukasz Cywi{\'n}ski}\email{cywinski@physics.ucsd.edu}
\author{Hanan Dery\footnote{Current address: Department of Electrical and Computer Engineering, University of Rochester, Rochester, NY 14627, USA}}
\author{Parin Dalal}
\author{L. J. Sham}
\affiliation{Department of Physics, University of California San Diego
La Jolla, California 92093-0319, USA}

\begin{abstract}
We treat the spin injection and extraction via a ferromagnetic
metal/semiconductor Schottky barrier as a quantum scattering problem.
This enables the theory to explain a number of phenomena involving
spin-dependent current through the Schottky barrier, especially the
counter-intuitive spin polarization direction in the semiconductor
due to current extraction seen in recent experiments. A possible
explanation of this phenomenon involves taking into account the
spin-dependent inelastic scattering via the bound states in the
interface region. The quantum-mechanical treatment of spin transport
through the interface is coupled with the semiclassical description
of transport in the adjoining media, in which we take into account
the in-plane spin diffusion along the interface in the planar
geometry used in experiments. The theory forms the basis of the
calculation of spin-dependent current flow in multi-terminal
systems, consisting of a semiconductor channel with many
ferromagnetic contacts attached, in which the spin accumulation
created by spin injection/extraction can be efficiently sensed by
electrical means. A three-terminal system can be used as a magnetic
memory cell with the bit of information encoded in the magnetization
of one of the contacts. Using five terminals we construct a
reprogrammable logic gate, in which the logic inputs and the
functionality are encoded in magnetizations of the four terminals,
while the current out of the fifth one gives a result of the
operation.
\end{abstract}
\keywords{spin injection/extraction; spin diffusion; spintronic devices.}
\maketitle

\section{Introduction}
{\flushleft Electrical} spin injection (transfer of spin
polarization by electrical current)  from a ferromagnet into a
paramagnet was first achieved in junctions between metals by Johnson
and Silsbee.\cite{Johnson_PRL85,Johnson_PRB88} The spin injection into semiconductors
has proven to be a harder task.\cite{Zutic_RMP04} In late 90s a
successful spin injection at low temperatures from Mn-doped diluted
ferromagnetic semiconductors\cite{Fiederling_Nature99,Ohno_Nature99,Oestreich_APL99,Jonker_PRB00}
gave new impetus to the field of semiconductor spintronics.
Injection from ferromagnetic metals at temperatures up to the room
temperature followed soon
afterwards.\cite{Zhu_PRL01,Hanbicki_APL02,Strand_PRL03}
Initially\cite{Zhu_PRL01} the injection had quite low efficiency,
which was was later
increased,\cite{Hanbicki_APL02,Hanbicki_APL03,Adelmann_PRB05} with
maximum reported\cite{Hanbicki_APL03} value of 30\%. The rise in
spin injection efficiency was achieved by a proper doping of the
metal/semiconductor
interface.\cite{Jonker_IEEE03,Adelmann_JVST05,Zega_PRL06} Spin
extraction was also seen through optical measurement of spin
accumulation in forward-biased MnAs/GaAs
junction.\cite{Stephens_PRL04}

During the last two years, there was a tremendous progress in both
spin injection and extraction in Fe/GaAs structures. The spin
accumulation due to both spin injection from Fe and spin extraction
from GaAs into Fe (a depletion of spins which can move more easily
into the magnet) were imaged by Kerr
spectroscopy.\cite{Crooker_Science05} Soon afterwards the spin
accumulation in the semiconductor near the junction with a magnet
has been sensed electrically,\cite{Lou_PRL06,Lou_NP07} proving that
the current through the metal/semiconductor junction depends on spin
polarization of electrons inside the semiconductor.

In these experiments\cite{Crooker_Science05,Lou_PRL06,Lou_NP07} an
unexpected sign of spin accumulation near the drain was seen. For
source and drain with parallel magnetizations, and for potential
drops at the interfaces small compared to the energy scale on which
the spin-projected densities of states in the ferromagnets change
significantly, spins of opposite directions should accumulate near
source and drain contacts. Contrary to this expectation, which is a
consequence of time-reversal symmetry for elastic tunneling between
the two materials,\cite{Imry} the observed spin accumulation near the drain had
the same sign as the one near the source contact. In
Sec.~\ref{sec:extraction} we review a theory\cite{Dery_PRL07} of
spin extraction which takes into account inelastic scattering
through the bound states near the interface. These arise from the
inhomogeneous profile of heavy n-type doping of the semiconductor
near the junction with the
metal.\cite{Jonker_IEEE03,Adelmann_JVST05,Zega_PRL06} This doping
makes the Schottky barrier thin enough  ($\sim$10 nm) for efficient
spin transport via tunneling. It also results in the creation of a
potential well for electrons next to the barrier. At
forward bias, the electrons from the bulk of the semiconductor
either tunnel directly into the metal (elastic process), or scatter
inelastically into the quasi-bound states in the well, and then leak
out into the magnet. These two channels of electron transport
through the interface favor opposite spin orientations, leading to
opposite signs of spin accumulation.
An alternative theory based on first-principles calculation of interface electronic structure (but neglecting the bending of the conduction band potential in GaAs) has also been proposed.\cite{Chantis}

Efficient spin injection/extraction is a basic prerequisite for any
kind of practical application of spintronics systems. The simplest
spintronics device is a two-terminal spin valve, in which a current
is passed between two ferromagnetic contacts connected through a
paramagnetic channel. However, in electronics three-terminal
semiconductor devices are indispensable for their switching (biasing
the gate of a field effect transistor) or amplification capabilities
(driving a current into the base of a bipolar transistor). Many
types of ``spin transistors'' have been proposed theoretically. The
most famous is the simple ``current modulator'' proposed by Datta
and Das\cite{DattaDas_APL90} in 1990, in which the electric field of
the gate together with Rashba spin-orbit interaction\cite{Bychkov_JPC84,Winkler} in the small-bandgap semiconductors
controls the spin precession of ballistic electrons injected and
extracted by ferromagnetic contacts. Despite a large experimental
effort a conclusive demonstration of the device operation has
remained elusive. Let us mention some of the other proposed
semiconductor spin-transistors. A diffusive version of a Datta-Das
system has been put forth.\cite{Schliemann_PRL03} Magnetic unipolar\cite{Flatte_APL01} and bipolar transistors\cite{Fabian_APL04,Fabian_PRB04,Zutic_JPC07} (both of which require
non-degenerate magnetic semiconductors) have been analyzed. Another
proposal was that of a spin transistor without any ferromagnetic
elements,\cite{Hall_APL03} which relies exclusively on strong
spin-orbit interaction experienced by electrons in small bandgap
materials such as InAs. There was also an idea of bypassing the
problems with efficient spin injection and using a proximity effect
of a ferromagnetic gate.\cite{Ciuti_APL02,McGuire_PRB04} Here, we
review our theoretical work on a class of systems consisting of a
semiconductor channel with multiple ferromagnetic contacts attached.
We work in the regime of diffusive spin transport  at room
temperature, and we concentrate on Fe/GaAs structures. However, we
do not exploit strong spin-orbit interaction present in III-V
semiconductors; the spin-orbit scattering serves only as a source of spin
relaxation. Consequently, the proposed devices are also suited for
silicon-based systems, as the spin relaxation time in Si is expected
to be at least an order of magnitude longer than in GaAs. This makes
silicon a perfect candidate for spintronics applications which do
not rely on spin manipulation through spin-orbit interaction.
The existence of spin-dependent coupling through a heavily doped
Schottky junction between Si and a paramagnetic metal has been shown
indirectly.\cite{Anderberg_PRB97} Recently, progress has been
made\cite{Min_NM06} in creating tunneling contacts with widely
tunable conductance between a ferromagnet and Si, albeit without
showing yet spin injection. It is encouraging that hot electron spin
injection into silicon accompanied by a magnetoresistive effect has
been achieved recently,\cite{Appelbaum_Nature07} as well as spin
injection from iron into silicon using aluminum oxide tunneling
barriers.\cite{Jonker_NP07}

Our review is organized as follows. In Sec.~\ref{sec:barrier} we
give a theory of spin-polarized transport through the heavily doped
metal/semiconductor interface. Spin injection at large reverse bias
is described in Sec.~\ref{sec:injection}. For spin extraction we
introduce in Sec.~\ref{sec:extraction} a new mechanism of spin
transport through the junction due to leakage of localized electrons
from the potential well close to the interface, and we show that it
gives an opposite sign of spin accumulation to the mechanism of
direct tunneling between the bulk of the semiconductor and the
metal. Then in Sec.~\ref{sec:low_bias} we consider a special case of
an ``optimally doped'' barrier (without a pronounced potential well)
kept at low bias, which can be used as an electrical probe of the
spin accumulation in the semiconductor. In
Sec.~\ref{sec:accumulation} we couple the description of spin
injection/extraction with the diffusive transport inside the
semiconductor, and we explain the significance of spin accumulation
for magnetoresistance of a two-terminal system (a spin valve). We
introduce the basic concept of electrical sensing of the spin
splitting in the semiconductor using a ferromagnetic contact in
Sec.~\ref{sec:multiterminal}. The possible applications, including
the Magnetic Contact Transistor\cite{Dery_MCT_PRB06} and a
magneto-logic gate integrable into a large-scale
circuit\cite{Dery_Nature07} are presented in
Sec.~\ref{sec:multiterminal} and are illustrated by theoretical
calculations.

\section{Spin transport through the Schottky barrier} \label{sec:barrier}
{\flushleft The} Schottky barrier\cite{Tunneling_Phenomena,Sze}
between a metal and a semiconductor is created by redistribution of
charges in the space charge layer. We denote the barrier height
measured from the Fermi level of the metals by $\phi_{B}$, and its
thickness as $d$. For uniformly n-doped semiconductor the barrier
shape is approximately parabolic, and the depletion width $d$  is
the distance between the interface and the onset of the bulk
flat-band region.

For Fe/GaAs junction\cite{Adelmann_JVST05} we will employ a value of
$\phi_{B}$$=$$0.8$ eV. For homogeneous doping $n_{0}$$<$$10^{17}$
cm$^{-3}$ we have the depletion width $d$$>$$100 $ nm. For such a
wide barrier the tunneling current is negligible, and
the current is due to purely classical thermionic emission\cite{Sze}
which depends on temperature and the barrier height, not on its width or
shape. Even if this current is spin polarized, its total density is
too small to create an appreciable spin accumulation. For higher
$n_{0}$ the tunneling dominates the transport through the junction,
but only for extremely high bulk doping levels
($n_{0}$$\sim$$10^{19}$ cm$^{-3}$) we have $d$$\approx$$10$ nm
resulting in  appreciable current densities.
In order to achieve such thin barriers yet with the bulk of the
semiconductor having the carrier density less than $10^{19}$
cm$^{-3}$, a strongly inhomogeneous doping profile has to be used
near the interface.\cite{Jonker_IEEE03,Adelmann_JVST05,Zega_PRL06}
Spin injection from Fe into GaAs with bulk $n_{0}$$\sim$$10^{16}$
cm$^{-3}$ has been
observed\cite{Hanbicki_APL02,Hanbicki_APL03,Crooker_Science05} only
in such heavily doped junctions, in which the first 15 nm of
semiconductor beneath the interface is doped with $n_{d}$$=$$5\cdot
10^{18}$ cm$^{-3}$ donors.

In general, the doping of the interface results in a creation of a
potential well close to the
barrier.\cite{Zachau_SSC86,Geraldo_JAP93,Shashkin_Semiconductors02}
Even if there is no well in equilibrium, at high forward bias when
less electrons need to be depleted from the semiconductor, the well
creation is inevitable. In Sec.~\ref{sec:extraction} we show that
the presence of bound states in this well can have a profound effect
on spin extraction from the semiconductor.

\subsection{Spin injection} \label{sec:injection}
{\flushleft Theoretical} analysis of spin injection from metals into
semiconductors has shown\cite{Schmidt_PRB00,Rashba_PRB00,Fert_PRB01}
that the junction with large resistance (a tunneling barrier) is
necessary for the current to be polarized. More precisely, since the
spin-depth conductance of the semiconductor
$G_{\text{sc}}$$=$$\sigma/L$ (with conductivity $\sigma$ and spin
diffusion length $L$) is much smaller than its metal counterpart
$G_{m}$$=$$\sigma_{m}/L_{m}$, for spin injection to occur the
junction conductance $G$ has to fulfill $G$$\leq$$G_{\text{sc}}$. In
such a case the spin polarization of the current at the interface is
determined by the spin-selectivity of the barrier, $\Delta G$$=$$G_{+}$$-$$G_{-}$, in which $G_{s}$
are the conductances for spin $s$$=$$\pm$ (along the quantization
axis given by magnetization of the ferromagnet). This ``conductivity
mismatch'' effect  was actually first analyzed in 1987  by Johnson
and Silsbee.\cite{Johnson_Silsbee_PRB87} From experiments on
Fe/GaAs, $\Delta G/G$$\leq$$0.3$ was deduced for spin
injection.\cite{Hanbicki_APL03} We also stress that although the
barrier with $G$$\ll$$G_{\text{sc}}$ gives spin-polarized currents,
the total current density can be too small to create an appreciable
spin accumulation in the
semiconductor.\cite{Fert_PRB01,Dery_lateral_PRB06,Fert_IEEE07} A
rule of a thumb is that $G$$\sim$$G_{\text{sc}}$ leads to efficient spin injection
(i.e.~resulting in large spin accumulation), but, strictly speaking, the
geometry of a system has to be taken into account when choosing the
optimal barrier parameters.\cite{Dery_lateral_PRB06}

An important quantity in the description of spin transport is a
spin-dependent electrochemical potential $\mu_{s}(\mathbf{x})$. It
is defined as
\begin{equation}
\mu_{s}(\mathbf{x}) = \mu^{c}_{s}(\mathbf{x}) -e\phi(\mathbf{x}) \,\, , \label{eq:mu_def}
\end{equation}
where $\mu^{c}_{s}$ is the chemical potential of electrons with spin $s$$=$$\pm$, $\phi$
is the electrostatic potential, and the elementary charge $e$$>$$0$.
The spin splitting of the electrochemical potential, $\Delta
\mu$$=$$\mu_{+}-\mu_{-}$, corresponds to the presence of
non-equilibrium spin density (spin accumulation). In a non-magnetic
material $\Delta \mu$$\neq$$0$ means $\Delta
n$$=n_{+}$$-$$n_{-}$$\neq$$0$, where $n_{s}$ is the density of
electrons of spin component $s$.

\begin{figure}[t]
\begin{center}
\includegraphics[width=8cm]{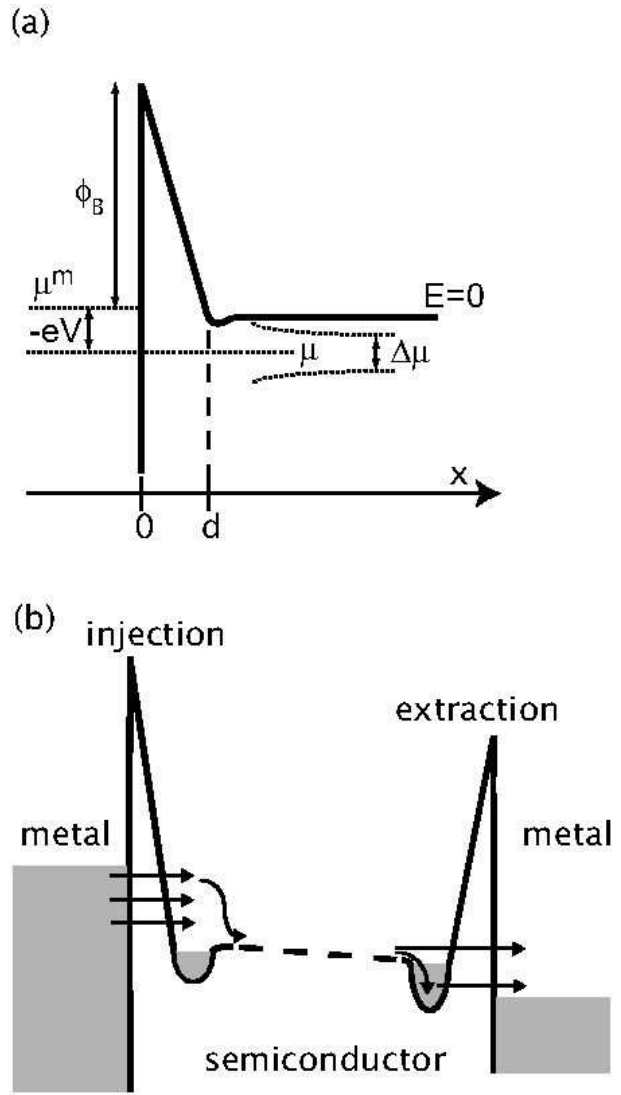}
\caption{(a) Heavily doped Schottky junction at reverse bias. The
semiconductor is non-degenerate (at room temperature). $\mu$ is the
average electrochemical potential at the onset of the flat-band
region, and $\Delta \mu$ is the spin accumulation. (b) Schematic
picture of current flow from the ferromagnetic injector down to the
ferromagnetic spin-extracting drain. The potential wells filled with
carriers near the interfaces are created by the inhomogeneous doping
profile (heavy $n^{+}$ doping near the interface). For extraction,
two routes are drawn: a direct tunneling from the bulk of the
semiconductor (elastic process) and tunneling of the electrons from the potential well into the magnet sustained by capture of bulk electrons (inelastic process). }
\label{fig:Schottky}
\end{center}
\end{figure}

In Fig.~\ref{fig:Schottky}a we show the energy diagram of the
Schottky barrier. We define the bias $eV$ applied to the junction as
the difference between the average electrochemical potential in the
flat-band region $\mu$$=$$(\mu_{+}+\mu_{-})/2$ and metal's
$\mu^{m}$. $V$$>$$0$ ($<$$0$) is forward (reverse) bias
corresponding to electrons going from (into) the semiconductor. This
definition of $V$ is convenient in the case of  spin accumulation small enough for
$\mu_{s}$ to be linearly proportional to the nonequilibrium parts of
the spin densities $\delta n_{s}$, see Sec.~\ref{sec:accumulation}.
Then, because of quasi-neutrality\cite{Smith_Semiconductors} we have
$\delta n_{+}+\delta n_{-}$$=$$0$ and $\mu$ is equal to the
equilibrium chemical potential in the semiconductor.

A schematic picture of spin injection and extraction through
Schottky barriers is shown in Fig.~\ref{fig:Schottky}b. The
junctions are much more resistive than the semiconductor channel, so
that the electrochemical potential shows discontinuities at the
barriers. Because of this and the large difference of conductivities
of the semiconductor and the ferromagnet, we can disregard both
spatial and spin dependence of $\mu_{s}$ in the ferromagnet, and use
a single value of chemical potential $\mu^{\text{m}}$. The current
injection occurs because of tunneling of electrons from the metal
into the semiconductor, described\cite{Tunneling_Phenomena} by
spin-dependent transmission coefficient of the particle flux
$T_{s}(k_{x},\mathbf{k}_{\parallel})$, with $k_{x}$ the wave-vector
in the direction of the interface, and $\mathbf{k}_{\parallel}$ the
in-plane wave-vector, which we assume conserved (specular
transmission).

We neglect the atomic structure of the Fe/GaAs
interface,\cite{Zega_PRL06,Chantis,Zwierzycki_PRB03} and use a
simplified band-structure for the bulk ferromagnet with a single
spin-split band with spin-dependent Fermi velocities $v_{m,s}$. For
iron, a model with effective mass $m_{m}$ equal to the free-electron
mass and Fermi wave-vectors $k^{m}_{+}$$=$$1.1$~\AA$^{-1}$ (
$k^{m}_{-}$$=$$0.42$~\AA$^{-1}$) for majority (minority) electrons
has been widely used.\cite{Slonczewski_PRB89,Ciuti_PRL02} Due to the
assumption of specular transmission only the electrons having
$v^{x}_{m,s}$$\approx$$v_{m,s}$ can tunnel into the $\Gamma$ valley
of the semiconductor's conduction band. We define the imaginary wave
vector within the barrier $\kappa$ and the corresponding velocity
$v_{b}$$=$$\hbar \kappa/m_{sc}$ (with semiconductor effective mass
$m_{sc}$). For high barrier and $m_{sc}$$\ll$$m_{m}$ considered here
we have $v_{b}$$\gg$$v_{m,s}$,$v_{sc}$, with the transmitted
electron velocity in the semiconductor $v_{sc}$. Within this model
we obtain for the flux transmission coefficient:
\begin{equation}
T_{s} \approx \frac{v_{m,s}v_{sc}}{v_{m,s}^{2} + v_{b}^{2}}
e^{-2\kappa d} \approx  \frac{v_{m,s}v_{sc}}{v_{b}^{2}}
e^{-2\kappa d} \equiv v_{sc} A_{s}e^{-2\kappa d}   \,\,
.\label{eq:T}
\end{equation}
If we approximate the barrier by a square step of thickness $d$,
then $\kappa$$=$$\sqrt{2m_{sc}(\phi_{B}+\mu-eV)}/\hbar$. For a
triangular barrier the expression for $\kappa$ has to be
modified,\cite{Osipov_PRB04} but the spin-dependent $A_{s}$ factor
remains the same. Consequently, electrons with larger velocity in
the metal tunnel more efficiently into the semiconductor. In iron
this translates into preferential injection of majority spins, which
is in agreement with experiments.\cite{Crooker_Science05} Within
this model we also expect that if Fe is replaced by a zinc-blende
MnAs, because of the different ratio between the majority and
minority spin wave-vectors,\cite{Zhao_PRB02} the spin-selectivity of
the junction will have opposite sign to the Fe case.

For large reverse bias  ($|eV|$$\gg$$k_{B}T$) the injected current
does not depend on the occupation function in the semiconductor,
since most of the electrons tunnel from the metal into the states at
least $k_{B}T$ above the chemical potential in the semiconductor
(see Fig.~\ref{fig:Schottky}b). Up to a certain critical reverse
bias the barrier thickness does not change much, only the well
becomes more shallow. Above this critical bias the electrons start
to be depleted from the bulk of the semiconductor. The wide
depletion region created then was shown to be detrimental to
spin injection.\cite{Albrecht_PRB03} Another reason for avoiding too
large reverse biases is that hot electron injection is accompanied
by enhanced spin relaxation in GaAs.\cite{Saikin_JPC06}

\subsection{Spin extraction in the presence of bound states near the interface} \label{sec:extraction}
{\flushleft An} analogous calculation of tunneling from the 3D
states in the bulk of the semiconductor into the metal (spin
extraction) gives the same spin selectivity, so that the spins
parallel to the minority spin in Fe should be accumulated, contrary
to the observation.\cite{Crooker_Science05} The experiments can be
explained by including the presence of electrons localized in in the
well near the interface, and considering a two-step process, in
which tunneling of the electrons from the bound states into the ferromagnet is followed by vacant states being filled by decay of extended state electrons (carrier capture). 

The spin-selectivity of the junction for free and localized
electrons is explained in the following way. For free electrons,
current conservation is well defined on both sides of the interface region
(elastic scattering). As explained before, in reverse or in low
forward bias the `effective velocity' in the barrier dominates and
the current scales with the electron's velocity in the metal side.
On the other hand, for localized electrons the conservation of total
reflection and transmission is irrelevant. Electrons escape from the
well into the vacant states in the ferromagnet and the transmitted
current scales with the decay rate of the bounded wave function. In
this case, the decay is fastest when the electron's velocity in the
ferromagnet matches the `effective velocity' in the well (being
inversely proportional to the well's width). Later we show, that in
the case of Fe/GaAs one gets an antipodal spin-behavior for free and
localized carriers.\cite{Dery_PRL07}

We denote the bulk doping density by $n_{0}$. The ultra-heavily
doped region at the junction has width $d$ and doping $n_{d}$. There
is also a transition region of width $d_{tr}$ where the donor
density interpolates between $n_{0}$ and $n_{d}$. The conditions for
the existence of the potential well are the
folllowing:\cite{Dery_PRL07}
\begin{eqnarray}
d & \approx & \sqrt{\frac{2\epsilon_{r}\epsilon_{0}\phi_{B}}{e^{2}n_{d}}} \,\, , \\
n_{d} & \gg & n_{0} \,\, , \\
d_{tr} & \sim & \lambda_{B} \,\, ,
\end{eqnarray}
where $\epsilon_{r}$ is the relative permittivity of the
semiconductor and $\lambda_{B}$ is the electron's de Broglie
wavelength (typically $\sim$$10$ nm in GaAs quantum wells). The
first equation determines  the Schottky barrier thickness $d$, the
second guarantees an excess of electrons in the transition region,
and the third one is for the two-dimensional character of the
electronic states in the well.

\begin{figure}[t]
\begin{center}
\includegraphics[width=9cm]{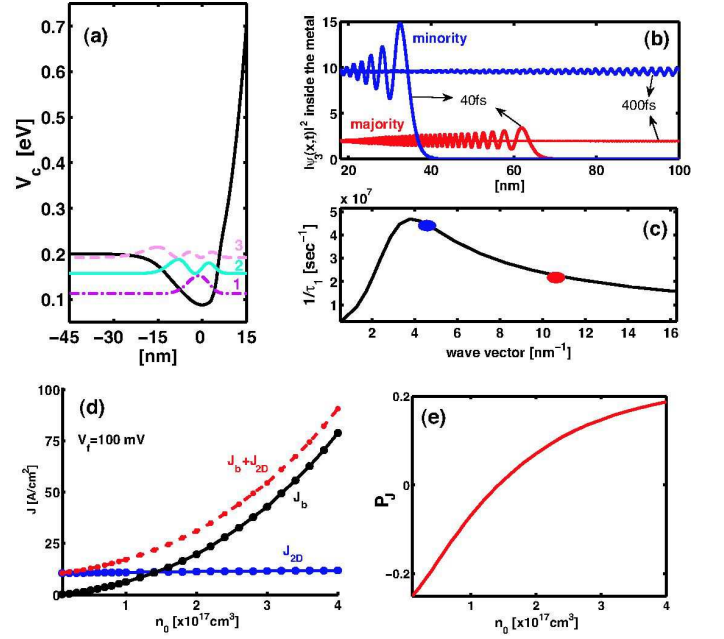}
\caption{(a) The self-consistent conduction band potential in the
semiconductor for $0.2$ V forward bias with the wave functions of
the three bound states. (b) Spin-dependent amplitudes of the wave
function penetrating into the ferromagnet  after 40 and 400 fs. (c)
Escape rate from the third bound state versus the wave-vector in the
metal. The upper (lower) mark refers to the minority (majority)
electrons in Fe. (d) Extracted current density as a function of the
backround doping $n_{0}$ at zero temperature. $J_{b}$ and $J_{2D}$ denote current due to free and localized electrons, respectively, and $J_{b}+J_{2D}$ denotes the sum of the two terms (the total current). (e) Spin polarization of the total current. Panels (d) and (e) are adapted from Ref.~20.} \label{fig:extraction}
\end{center}
\end{figure}

In Fig.~\ref{fig:extraction}a we show the results of the
self-consistent calculation of Schr{\"o}dringer and Poisson
equations for the conduction band profile near the junction at $0.2$
V forward bias using  $n_{0}$$=$$3.6\cdot 10^{16}$ cm$^{-3}$,
$n_{d}$$=$$5\cdot 10^{18}$ cm$^{-3}$, $d$$=$$d_{tr}$$=$$15$ nm and
$\epsilon_{r}$$=$$12.6$. These parameters are chosen to approximate
the junctions used in the
experiments.\cite{Crooker_Science05,Lou_PRL06,Lou_NP07} At this bias
three bound states are present in the well.

The process of electron escape from the well into the continuum in
the ferromagnet is calculated in the following way.\cite{Dery_PRL07}
At time $t$$=$$0$ the wave functions for spin $s$$=$$\pm$ are
identical, taken as the $i$th bound state, and they are equal to
zero on the metal side. The time-dependent Schr{\"o}dringer
equations is then solved numerically using the potential from
Fig.~\ref{fig:extraction}a. Inside the metal (iron) we use the same
simplified model as before.
For the calculation we have used a one-dimensional box of 200 nm
width encompassing both the well and the inside of the metal, and we
have used discrete transparent boundary conditions to prevent
reflections from the edges.\cite{Arnold_CMS03}
Fig.~\ref{fig:extraction}b shows the wave functions of the third
quasibound state penetrating into the metal at times $t$$=$$40$ and
$400$ fs. The electrons with minority spin in Fe have bigger
penetration amplitude, and this behavior persists for longer times
and for all the bound states. The escape rate is practically
constant in time (resulting in exponential decay of the quasibound
state) and is given by
\begin{equation}
\frac{1}{\tau^{esc}_{i,s}} = - \frac{1}{\int_{box} dx
|\psi_{i,s}(x,t)|^{2}} \frac{d}{dt} \int_{semiconductor}
\!\!\!\!\!\!\!\!\!\!\!dx |\psi_{i,s}(x,t)|^{2} \,\, .
\end{equation}
Fig.~\ref{fig:extraction}c shows the escape rate from the first
quasi-bound state versus the wave-vector in the metal $k_{m}$. The
escape rate has maximum when the `effective velocity' in the well
matches the velocity in the metal:
$k_{m}/m_{m}$$\sim$$\pi/(m_{sc}d_{tr})$. The values of Fermi
wave-vectors which we use for iron are on the right side of the
curve, where the escape rate decreases with $k_{m}$. Such behavior
for high metal wave-vectors agrees with the extended WKB model for
the alpha-particle decay,\cite{Kemble} in which the coupling between
a quasi-bound state and the continuum scales with the inverse
wave-vector in the continuum.

The spin-dependent current density $J_{2D,s}$ due to leakage of
localized electrons is proportional to the areal electron density in
the $i$th state $\tilde{n}_{i,s}$ (with energy above the metal's
Fermi level)  divided by the escape time: $J_{2D,s}$$\propto$$
\sum_{i} \tilde{n}_{i,s}/\tau^{esc}_{i,s}$. The spin relaxation time
in the well is of the order of tens of ps,\cite{Malinowski_PRB00}
whereas the escape time is $\sim$1 ns, so that we have little spin
accumulation in the well: $\tilde{n}_{i,s}$$\simeq$$\tilde{n}_{i}/2$.
The electron which escapes from the well into the magnet is
replenished by an electron with the same spin from the bulk region
due to spin-conserving capture process of free electrons by the
well, e.g. by emission of longitudinal optical phonons or
carrier-carrier scattering with degenerate electrons in the well.
These processes are much faster than the spin relaxation time in the
well.\cite{Deveaud_APL88} Thus, the bulk region is left with
more spin-up (down) electrons if it provides the well with more
spin-down (up) electrons. The necessity of the capture process is
consistent with the longitudinal optical phonon signature in the low
temperature conductance measurement of Fe/GaAs by Hanbicki et
al.\cite{Hanbicki_APL03} The presence of this signature for the
forward bias had been an open question, and our model suggests a
possible explanation.

Apart from the current $J_{2D}$ due to the escape from the well,
there is always some current $J_{b}$ due to direct tunneling between
the bulk of the semiconductor and the metal.
Fig.~\ref{fig:extraction}d shows the contributions to the total
current due to the elastic and inelastic processes versus the
bulk doping $n_{0}$, calculated at zero temperature. The
transmission from the bulk states is calculated for the
self-consistent potential using the transfer matrix method,
including the resonant behavior of free electrons due to the well
potential.\cite{Stiles_PRB93} Since this potential is weakly
affected by changing $n_{0}$ as long as $n_{0}$$\ll$$n_{d}$, the
$J_{2D}$ current increases only by 10\% in the shown range of
$n_{0}$. On the other hand, $J_{b}$ depends strongly on $n_{0}$: as
the chemical potential in the bulk increases with $n_{0}$, the
number of carriers which can tunnel into the metal increases.
Fig.~\ref{fig:extraction}e shows the spin polarization of the total
current $P_{J}$$=$$(J_{+}-J_{-})/J$. The critical background doping
at which $P_{J}$$=$$0$ is $1.5\cdot 10^{17}$ cm$^{-3}$. This is not
exactly the density at which $J_{b}$$=$$J_{2D}$ because
$|P_{J_{b}}|$ and $|P_{J_{2D}}|$ are slightly different.

Because the effective spin selectivity of the junction depends on
the carrier density and the bending of the conduction band, it can be
controlled by voltage in the electrical spin
switch.\cite{Dery_PRL07} In a semiconductor channel of thickness
$h$$\sim$$100$ nm, a voltage applied to a back-gate deposited on the
opposite side of the channel to the ferromagnetic contact can be
used to manipulate the density of free electrons below the magnet.
By varying this back-gate voltage $V_{G}$ in a properly designed
structure it is possible to switch between the transport regimes
dominated by tunneling from the bulk and escape from the well, and
thus change the sign of the spin accumulation in the
semiconductor.\cite{Dery_PRL07}

\subsection{Spin-dependent conductance at low bias} \label{sec:low_bias}
{\flushleft Now} let us analyze the conductance of the interface with a
very shallow well at small bias (an ``optimally doped'' barrier). We
concentrate on the room-temperature case relevant for potential
applications, and take the bulk of semiconductor as non-degenerate.
An example of a doping profile yielding such an interface is the
$\delta$-doping,\cite{Zachau_SSC86,Geraldo_JAP93,Shashkin_Semiconductors02}
in which a single monolayer of a semiconductor material near the
interface is doped with a donor density impossible to achieve in the
bulk material. When a $\delta$-doping layer is placed at a distance
$d_{0}$ from the interface, and its planar density is
$n_{\text{2D}}$$=$$ \epsilon_r\epsilon_{0}(\phi_{B}+\mu)/4\pi$$e^{2}d_{0}$,
then the barrier shape is triangular, there is no well at zero bias,
and the barrier width is $d_{0}$. For $d_{0}$$=$$5$ nm,
$\phi_{B}$$=$$0.8$ eV, and $\mu$$=$$-0.1$ eV (corresponding to
bulk $n$$=$$10^{16}$ cm$^{-3}$), we get
$n^{0}_{\text{2D}}$$\approx$$10^{13}$ cm$^{-2}$.

For the ``optimally doped'' contact only the elastic transport
channel is present, and we can derive an
expression\cite{Osipov_PRB04} for the current with spin
$s$.
In the regime of bias $|eV|$, $|\mu^{m}-\mu_{s}|$, and
$|\mu_{s}-\mu|$ smaller than $k_{B}T$ the formulas for
$j_{s}(V,\mu_{s})$ can be linearized around $V$$=$$0$, and we obtain
for the spin currents:
\begin{eqnarray}
j_{s} & \approx & \frac{G_{s}}{e} (\mu^{m}_{s}-\mu_{s} ) \,\, ,\label{eq:Schottky_boundary} \\
G_{s} & = & \frac{4e^{2}}{m_{\text{sc}}} A_{s} e^{-2\kappa d} \,\,
n_{0} \,\, , \label{eq:Gs}
\end{eqnarray}
where $G_{s}$ is the barrier conductance at low bias, the
spin-dependence of which comes from the $A_{s}$ factor. As 
discussed above, the ratio $G_{+}/G_{-}$ is equal to the ratio of
the velocities of carriers with different spin in the ferromagnet,
which is approximately 2 in our model in agreement with spin-LED
experiments.\cite{Hanbicki_APL02,Hanbicki_APL03} For
$n_{0}$$=$$10^{16}$ cm$^{-3}$ and $d_{0}$$\approx$$5$ nm we obtain
$G_{s}$ of the order of $10^{3}$ $\Omega^{-1}$cm$^{-2}$.
In the following sections, we will use the above model of
spin-dependent properties of the junction  to model how the spin
accumulation can be sensed electrically by a contact kept at low
bias.

\section{Spin accumulation in the diffusive spin valve} \label{sec:accumulation}
{\flushleft We} work in the diffusive regime, in which the spin
relaxation time $\tau_{sr}$ is much longer than the momentum
scattering time. Then, from the Boltzmann equation we can derive the
spin diffusion
equation\cite{Valet_PRB93,Hershfield_PRB97,Villegas_JAP07,Cywinski_thesis}
for the non-equilibrium parts of the spin densities $\delta n_{s}$.
In a paramagnetic and non-degenerate semiconductor the
electrochemical potential $\mu_{s}$ defined in Eq.~(\ref{eq:mu_def})
is given by
\begin{equation}
\mu_{s}  =  k_{B}T \ln \Big( \frac{n_{0}/2+\delta
n_{s}}{n_{0}/2}\Big) -e\phi \,\, \simeq \,\, k_{B}T\frac{\delta
n_{s}}{n_{0}/2} - e\phi \,\, ,    \label{eq:mu}
\end{equation}
where $n_{0}$ is the total carrier density, and the second
expression is the linear approximation valid for
$|\Delta\mu|$$<$$k_{B}T$ (equivalently $|\delta
n_{s}|$$<$$n_{0}/2$). Here we concentrate on the linear regime and
low electric fields,\cite{Yu_Flatte_long_PRB02} so that we can write
the diffusion equation  for the spin-splitting of the
electrochemical potentials:
\begin{equation}
\nabla^{2}  \mu_{s} = \frac{\mu_{s}-\mu_{-s}}{2L^{2}} \,\, ,  \label{eq:diffusion}
\end{equation}
where the spin diffusion length is defined in terms of diffusion
constant $D$ and spin relaxation time by $L$$=$$\sqrt{D\tau_{sr}}$.
For the spin-$s$ current in the paramagnetic semiconductor we have
\begin{equation}
\mathbf{j}_{s} = \frac{\sigma_{s}}{e}\nabla \mu_{s} =
\sigma_{s}\mathbf{E} + eD\nabla n_{s} \,\, ,
\end{equation}
where the conductivity for spin $s$ is $\sigma_{s}$$=$$n_{s}e\nu$
with $\nu$ being the mobility, and $\sigma_{+}$$\simeq$$\sigma_{-}$
to the first approximation in the linear regime. The only way for
the semiconductor to support a non-zero spin polarization of the
current is by creating a net spin density ($n_{+}$$\neq$$n_{-}$
corresponding to $\Delta \mu$$\neq$$0$).

In the spin valve the current is passed through a paramagnetic
channel between two ferromagnetic contacts. We assume collinear
magnetizations, oriented either parallel (P) or antiparallel (AP)
with respect to each other. If the distance between the contacts is
smaller than spin diffusion length $L$, the spin accumulation in the
channel depends on the alignment of the magnets. Provided that the
spin transport mechanism is the same for injection and extraction,
in P configuration the same spin species is preferentially injected and
extracted. The spin accumulation has opposite signs in the
neighborhood of the two contacts, and by $|\Delta \mu^{P}|$ we
denote the magnitude of spin splitting near the junctions. In the AP
configuration, spins of opposite directions are more easily injected
and extracted, resulting in large and nearly uniform spin
accumulation: $|\Delta \mu^{AP}|$$\gg$$|\Delta \mu^{P}|$. If the
mechanisms of spin injection and extraction are different, and the
spin selectivities of injecting and extracting junctions are
opposite as described in Sec.~\ref{sec:extraction}, the labels P and
AP refer not to the relative orientations of contact magnetizations,
but to the spin accumulation patterns described above.

The magneto-resistive (MR) coefficient of the spin valve is commonly
defined as $MR$$\equiv$$(I^{P}$$-$$I^{AP})/I^{P}$, where $I^{P(AP)}$
is the total current between the two terminals. A qualitative
relation between the MR and the spin accumulation can be derived
using the simple boundary conditions for currents from
Eq.~(\ref{eq:Gs}).

For the channel with relevant dimensions smaller than L, by
balancing the net spin injection into the channel with the spin
relaxation we obtain
\begin{equation}
MR = \frac{\Delta G}{G} \frac{\Delta \mu^{AP}}{eV} =
\Big(\frac{\Delta G}{G}\Big)^2 \Big(
1+\frac{\mathcal{V}}{2\mathcal{A}L}\frac{G_{sc}}{G} \Big)^{-1} \,\,
,
\end{equation}
where $\mathcal{V}$ is the volume of the channel and $\mathcal{A}$
is the area of the contacts, which are assumed to both have the same
$G_{s}$ and $\mathcal{A}$ for simplicity. The MR depends on the
ratio of $\Delta \mu^{AP}$ to the applied bias, which for small
electric fields is just a geometry-dependent constant. For realistic
parameters of a Fe/GaAs spin valve we obtain $MR$$\ll$$(\Delta
G/G)^{2}$, and a typical value of MR is about 1\% after optimizing
the system's geometry.\cite{Dery_lateral_PRB06} In a one-dimensional
channel of length $l$ we have $MR$$\sim$$(G/G_{sc})$$\cdot$$(L/l)$ which is quite small for realistic values of $G$ and practical values of $l$$\sim$$100$ nm. On the other hand the ratio of spin splittings $|\Delta \mu^{AP}/\Delta
\mu^{P}|$$\sim$$(2L/l)^{2}$ can be quite large even when $MR$ is small. This large difference of spin accumulations in P and AP cases is not expressed in a two-terminal system in the most effective way by the magnetoresistive effect.

The MR of a one-dimensional spin valve has been calculated
analytically\cite{Fert_PRB01} also using the boundary conditions
from Eq.~(\ref{eq:Gs}). The spin valve in the lateral geometry
relevant for experiments\cite{Crooker_Science05,Lou_PRL06,Lou_NP07}
has been analyzed qualitatively\cite{Fert_PRB01,Fert_IEEE07} and
quantitatively.\cite{Dery_lateral_PRB06} In the latter work an
effective one-dimensional diffusion equation was derived, accurately
describing the spin diffusion in a layer of material of thickness
$h$ smaller than the spin diffusion length $L$, and covered by
contacts with junction conductances $G$ smaller than the conductance
$\sigma/h$ of the underlying semiconductor layer. In a structure
like the one shown in Fig.~\ref{fig:MCT}a, we calculate the spin
diffusion by introducing the layer-averaged electrochemical
potential $\xi_{s}(x)$ in the semiconductor channel:
$\xi_{s}$$=$$\frac{1}{h}\int^{h}_{0}dy \, \mu_{s}(x,y)$. By
integrating out the $y$ dependence from Eq.~(\ref{eq:diffusion}) we
obtain the approximate equation:
\begin{equation}
\frac{\partial^{2}\xi_{s}}{\partial x^{2}} =
\frac{\xi_{s}-\xi_{-s}}{2L^{2}} + \frac{2G_{s}}{\sigma h}( \xi_{s} -
\mu^{m} ) \,\, ,
\end{equation}
where the second term on the right-hand side is present only under
the contacts, and $\mu^{m}$ is the electrochemical potential in the
ferromagnet. This equation is derived using the boundary condition
from Eq.~(\ref{eq:Gs}), and assuming small electric fields and small
spin accumulations (so that $\sigma_{s}$$\approx$$\sigma/2$). For
Fe/GaAs structures with $G_{s}$$\sim$$10^{3}$ $\Omega^{-1}$cm$^{-2}$
this approximate formalism gives results indistinguishable from
exact numerical calculations, and all the results presented below
are obtained using this approach.

\section{Electrical expression of spin accumulation in multi-terminal systems} \label{sec:multiterminal}
{\flushleft In} the previous section we have seen that in the spin
valve the patterns of spin accumulation in the semiconductor are
qualitatively different for P and AP configurations, but the MR
ratio does not directly reflect this feature. In order to achieve a
more efficient electrical expression of spin accumulation one has to
move  beyond a passive two-terminal device such as spin valve, and
consider a spin-transistor system in which additional external
stimuli can control the magnetoresistive effects. Below we review
several proposals of devices consisting of more than two
ferromagnetic terminals connected to a semiconductor channel. Their
common feature is the use of a ferromagnetic contact kept close to
zero bias (Sec.~\ref{sec:low_bias}), which is used to sense the spin
accumulation in the semiconductor beneath it.

\subsection{Magnetic Contact Transistor} \label{sec:MCT}
\begin{figure}[t]
\begin{center}
\includegraphics[width=8cm]{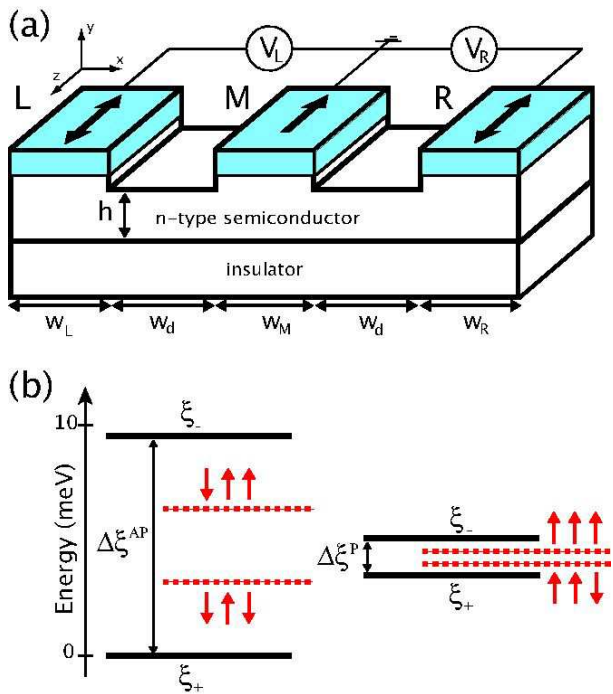}
\caption{(a) The three-terminal Magnetic Contact Transistor (MCT).
The outer edges if the channel are removed in order to confine the
spin accumulation under and between the
contacts.\protect\cite{Dery_lateral_PRB06} The R contact is
connected either to a controllable voltage $V_{R}$, or to a
capacitor C which maintains zero steady state current in this
terminal.  (b) A schematic picture illustrating the principle of the
MCT operation. The solid lines are the electrochemical potentials
($\xi_{s}$) in the channel beneath the R contact for AP and P
alignments of the L and M terminals. The dashed lines show the values of
the electrochemical potential inside the R contact for which the
$I_{R}$ current is quenched for a given alignment of all
magnetizations (represented by three arrows). The energy scale and
{\protect $\Delta \xi^{AP}/\Delta \xi^{P}$} ratio correspond to an
MCT with barrier conductances $G$$=$$10^{4}$ $\Omega^{-1}$cm$^{-3}$,
dimensions {\protect $w_{L}$$=$$w_{M}$$=$$w_{R}$$=$$400$} nm,
{\protect $w_{d}$$=$$200$} nm, {\protect $h$$=$$100$} nm, and
{\protect $V_{L}$$=$$0.1$ V.} } \label{fig:MCT}
\end{center}
\end{figure}

{\flushleft The} Magnetic Contact Transistor (MCT) consists of three
ferromagnetic contacts deposited on top of the paramagnetic channel
(see Fig.~\ref{fig:MCT}a). We concentrate on situation in which most of the current driven by the voltage $V_{L}$ passes between the
left (L) and middle (M) contacts, but other arrangements are possible.\cite{Saha_APL07} The voltage $V_{R}$ is adjusted to keep the R junction at low
bias,\cite{Dery_MCT_PRB06} and we describe its spin-dependent
conductance using Eq.~(\ref{eq:Gs}).
Alternatively,\cite{Cywinski_APL06} the R terminal can be connected
to a capacitor $C$, which adjusts the voltage of the R terminal so
that there is no net charge current in the steady state. We will
refer to P (AP) configurations of the L and M magnets as corresponding
to the spin accumulation patterns described before. In this section
we keep the M magnet fixed, and consider the P and AP alignments of
the L magnet with respect to M.

Let us first consider a situation in which the R terminal is
inactive. Then the L and M contacts constitute a spin valve with
voltage $V_{L}$ applied to it. For P and AP alignments of L and M
the $I_{L}$ current  is practically the same, but the spin
splitting of layer-averaged electrochemical potential in the channel
$\Delta \xi$ varies between two very different values. Beneath the
injecting and extracting contacts we have $|\Delta\xi^{AP}/\Delta
\xi^{P}|$$\propto$$(2L/l)^2$, with the effective length of the active
channel covered by L and M terminals
$l$$\approx$$w_{d}$$+$$w_{L}$$+$$w_{M}$. The spins accumulated
beneath the M terminal diffuse out to the right, but if
$w_{d}$$\ll$$L$ the spin accumulation beneath the R magnet is
practically the same as beneath the M contact.

The operational principle of the ``static'' MCT relies on the fact
that we actively control the $V_{R}$ voltage.\cite{Dery_MCT_PRB06}
In either P or AP alignment we bias the R terminal so that
$I_{R}$$=$$0$. If we then flip the L magnetization, {\it a finite
$I_{R}$ current will flow}. This is a consequence of large spin
accumulation in AP configuration, and spin selectivity of the R
junction ($\Delta G_{R}$$\neq$$0$). Using $|\Delta \xi^{P}|$$\ll$$
\Big| \Delta \xi^{AP} \Big| $ the ``on'' current is given by
\begin{equation}
|I^{on}_{R}| \approx \Big| \frac{\Delta G_{R}}{e} \frac{\Delta
\xi^{AP}}{2} \Big| \mathcal{A} \,\, .  \label{eq:Ion}
\end{equation}
Thus, we have found a way to {\it digitize the MR effect in the R
contact}. Instead of some finite ratio of P and AP currents, we can
have zero current for one and a finite current for the other
configuration. Even after taking the voltage noise in the system into account,
the ``on'' and ``off' currents should be easily discernible for
realistic parameters of Fe/GaAs system.\cite{Dery_MCT_PRB06} This digitization of the magnetoresistance had been observed recently in MnAs/GaAs three-terminal structure.\cite{Saha_APL07}

The digitization holds for the MR effect measured in the third
(R) terminal. The  larger currents in the other two (L and M) have a
small relative change when we alternate between P and AP
configurations. Yet these contacts do almost all of the job of
injecting and extracting spin-polarized currents. We can say that we
have transferred the magneto-resistive effect to the third contact,
where we can tune it by $V_{R}$ voltage. We have called this {\it
spin transference}.\cite{Dery_MCT_PRB06}  An alternative term of
``transferable magnetoresistance''  underlines the connection to
standard transistors.

To sketch the derivation of the above effect we write the spin
dependent electrochemical potentials underneath the R terminal as
\begin{equation}
\xi_{\pm} = \xi \pm \frac{1}{2}\Delta \xi \,\, ,
\end{equation}
where $\xi$ is the average value of electrochemical potential. If we
apply any voltage $V_{R}$, the situation inside the channel will
change in general. However, we are interested in $V_{R}$ such that
there is only a small (possibly zero) $I_{R}$ current. Then the spin
accumulation determined by the larger $I_{L}$ injected into the
channel remains practically unaffected. For such a voltage applied
to R we have the current density entering the R contact:
\begin{equation}
j_{R} = \frac{G_{R}}{e} (\mu_{R} - \xi) - \frac{\Delta G_{R}}{e}
\frac{\Delta \xi}{2} \,\, ,  \label{eq:jR}
\end{equation}
where $\mu_{R}$$=$$-eV_{R}$, $G_{R}$ is the total conductance of the
R junction, and $\Delta G_{R}$$=$$G^{R}_{+}-G^{R}_{-}$ is its spin
selectivity. The corresponding spin current  density $\Delta
j$$=$$j_{+}-j_{-}$ is
\begin{equation}
\Delta j_{R} = \frac{\Delta G_{R}}{e} (\mu_{R} - \xi) -
\frac{G_{R}}{e} \frac{\Delta \xi}{2} \,\, .
\end{equation}
We denote by $\mu_{0}$ the electrochemical potential in R
magnet that quenches the total current in this contact:
\begin{equation}
\mu_{0} = \xi + \frac{\Delta G_{R}}{G_{R}} \frac{\Delta \xi}{2} \,\, . \label{eq:muR}
\end{equation}
Plugging this $\mu_{0}$ for one L/M alignment into Eq.~(\ref{eq:jR}) with $\Delta \xi$ corresponding to the other L/M alignment we obtain Eq.~(\ref{eq:Ion}).
Depending on the alignment of the L and R magnets, the $V_{R}$ voltage
which corresponds to $I_{R}$$=$$0$ takes on four possible values
denoted by dashed lines in Fig.~\ref{fig:MCT}b. When the R terminal
is connected to the capacitor, we have four possible charge states
of the capacitor in the steady state. We discuss the transient
currents driven by L or R magnetization dynamics for this case in
Sec.~\ref{sec:dynamical} below.

In the ``off'' state, when $V_{R}$ is adjusted so that the net
$I_{R}$$=$$0$, there is a non-zero spin current flowing into the R
terminal:
\begin{equation}
\Delta j^{off}_{R} = \frac{\Delta \xi^{P,AP}}{2e}\frac{ \Delta
G_{R}^{2} - G_{R}^{2}}{G_{R}} \,\, . \label{eq:spin_current}
\end{equation}
The electrons with opposite spins flow in the opposite directions,
giving zero charge current, but adding to a net flow of spin
polarization. A similar effect was predicted in a lateral structure
with non-magnetic source and drain and two ferromagnetic gates, into
which the current could leak.\cite{McGuire_PRB04} A finite spin
current entering a ferromagnet can lead to reversal or
precession of the magnetization due to spin-transfer torque if the
polarization of injected spins is non-collinear with the
magnetization axis of the magnet.\cite{Stiles_TAP06} This effect has
been observed in an all-metallic system,\cite{Kimura_PRL06} in which
the magnetization of a floating terminal (zero net charge current
through it) was switched by a pure spin current. However, in our
case the relatively low carrier density in the semiconductor
together with the resistive contacts cannot transfer enough angular
momentum for such switching to occur.

A system similar to the MCT has been known for some time in the
field of all-metal magnetoelectronics as a non-local spin
valve.\cite{Johnson_Science93,Jedema_Nature01,Ji_APL06} However, the
third contact in the non-local spin valve is used as a passive
floating terminal (essentially a spin dependent voltage probe). In
the MCT all the contacts are active terminals controlled by applied
voltages. The possibility of control is closely related to the use
of the non-degenerate semiconductor as a paramagnetic channel. Due
to very small concentration of carriers compared to metals, spin
injection can lead to spin splittings of electrochemical potentials
of the order of milivolts in the channel. Then, voltages
supplied with mV accuracy can efficiently tune the magnetoresitive
effect measured in one of the terminals.

\subsection{Electric readout of magnetization dynamics} \label{sec:dynamical}
\begin{figure}[t]
\begin{center}
\includegraphics[width=8cm]{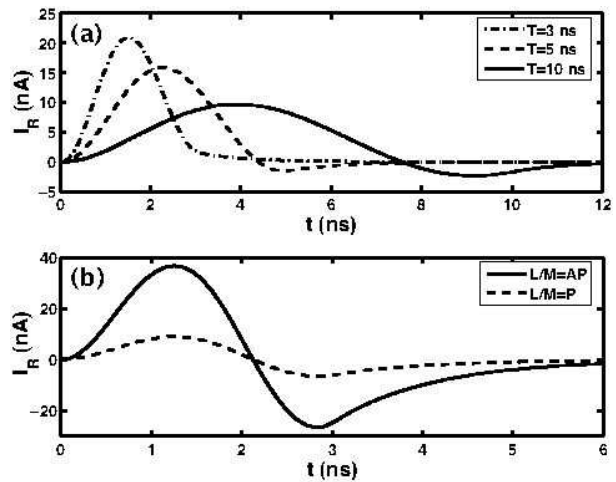}
\caption{(a) R current signal for reversal of L magnetization
occurring on time-scale of 3, 5 and 10 ns starting from AP alignment
of L relative to M magnet. (b) R current signal for $2\pi$ rotation
of R magnet for P and AP alignments of L and M magnets. The period
of rotation is 3 ns. The conductance of the barriers $G$$=$$10^{4}$
$\Omega^{1}$cm$^{-2}$. The area of a junction is 1 $\mu$m$^{2}$, and
the barrier thickness is taken to be 10 nm, resulting in junction
resisitance $R_{B}$$=$$10$ k$\Omega$ and capacitance $C_{B}$$=$$10$
fF. The external capacitance is $C$$=$$40$ fF. The channel is GaAs
at room temperature, with carrier density
$n_{0}$$=$$10^{16}$cm$^{-3}$. Adapted from Ref.~69. } \label{fig:capacitor_current}
\end{center}
\end{figure}

The MCT can also be used for electric measurement of magnetization
dynamics and dynamical readout of magnetization alignment. With the
R contact connected to a capacitor C there is zero $I_{R}$ current
in the steady state.\cite{Cywinski_APL06} For any alignment of all
the magnetizations, the charge on the capacitor adjusts itself so
that the electrochemical potential in the R contact is equal to the
current-quenching $\mu_{0}$, and there is no need for external
voltage tuning. When either the $L$ or the $R$ magnet starts to reverse due
to an application of an external magnetic field pulse, the potential in
the $R$ contact changes between two steady-state values (dashed
lines in Fig.~\ref{fig:MCT}) corresponding to the initial and final
alignments of the magnetizations. The measurement of the accompanying
transient $I_{R}(t)$ current recharging the capacitor allows for
electrical monitoring of magnetization dynamics. Alternative
application is a dynamical readout of the L/M alignment. In the P
case, the $2\pi$ rotation of the R magnet results in the
$I_{R}(t)$ current oscillation of much smaller amplitude than for
the AP case (see Fig.~\ref{fig:capacitor_current}).

The on-chip manipulation of L and R magnetizations is possible using
the architecture\cite{Prinz_Science98,Tehrani_IEEE03} of Magnetic
Random Access Memory (MRAM). In MRAM the nanomagnets (with typical
size similar to what we use in our modeling) arranged in a square
array are addressed using a network of current-carrying wires
positioned above and below the magnets. Pulses of current generate
time-dependent magnetic fields through Ampere's law, and these field
can be used to switch each of the magnets separately.

The capacitor $RC$ time (with $R$ being of the order of the junction
resistance, as they are the most resistive elements
in the circuit) has to be at most comparable to the magnetization
dynamics time-scale. For $G$$\sim$$10^{4}$ $\Omega^{-1}$cm$^{-2}$
and junction area of one $\mu$m$^{2}$, the capacitance $C$$=$$40$ fF
used below results in $RC$ time of about a nanosecond. In
all-metallic systems, in which the junctions are much less
resistive, the $RC$ time is not a problem. However, typical spin
accumulation in paramagnetic metal corresponds to $\Delta \mu$ (and
consequently the voltage swing on the capacitor) of  less than
$\mu$V.\cite{Jedema_Nature01,Ji_APL06} Then, for the transient
current to be measurable one has to use a nearly macroscopic
capacitor, which rules out an application in integrated circuits.
Again, the small carrier density in a semiconductor (allowing for
$\Delta \mu$$\sim$$10$ mV) is indispensable.

In order to model the transient behavior we add the time dependence
to the formalism of lateral spin diffusion. We are  interested in
time-scales of a least tens of picoseconds. The fastest
magnetization reversal time is about 100 ps,\cite{Gerrits_Nature02}
and magnetization switching times used in commercial devices are of
the order of a nanosecond. Thus, we use an adiabatic approximation
with respect to the processes occurring on a much shorter
(sub-picosecond) time scales: the  momentum scattering and
dielectric relaxation.\cite{Smith_Semiconductors}

The time-dependent diffusion equation for spin splitting of the
layer-averaged electrochemical potential $\Delta \xi$ is
\begin{equation}
\frac{\partial \Delta \xi}{\partial t} = D \frac{ \partial^{2}
\Delta \xi}{\partial x^{2}} +
\frac{\beta_{i}(t)}{\tau_{sr}}(\mu^{m}_{i}-\xi)
-\frac{\alpha_{i}}{2\tau_{sr}} \Delta \xi - \frac{\Delta
\xi}{\tau_{sr}} \,\,. \label{eq:Dxi}
\end{equation}
The $\alpha$($\beta$) dimensionless parameter is given by
$2L^{2}(G_{+}$$+$$(-)G_{-})/\sigma h$. The dynamics of magnetization
is parametrized by $\beta(t)$$\sim$$\Delta G(t)$, which
characterizes the contact polarization only along the $z$ axis. If
we deal with a coherent precession of magnetization then this is an
approximation. In principle, one should treat the diffusion of spin
accumulation treated as a vector quantity,\cite{Saikin_JPC04} and
take into account the noncollinearity of spins and the magnets in
the tunneling
process.\cite{Ciuti_PRL02,Brataas_PRL00,Brataas_EPJB01} However, for
tunneling barriers the non-trivial effects of this noncollinearity
are expected to be small,\cite{Brataas_PRL00,Brataas_EPJB01} and the
only thing that matters is the average polarization along the $z$
direction. Then we can model the influence of the contact with
magnetization making an angle $\theta$ with the $z$ axis by assuming
that $\beta$$\sim$$\Delta G\cos\theta$. On the other hand, if the
magnetization reversal is incoherent (e.g. proceeding by nucleation
of domains with opposite magnetization), the parameter $\beta(t)$
describes an area average of spin-selectivity of magnetically
inhomogeneous contact, and it is proportional to the $z$
component of the contact's magnetization.

The dielectric relaxation (about 100 fs for non-degenerate
semiconductor with $n_{0}$$=$$10^{16}$ cm$^{-3}$) is much faster
than the time-scale of magnetization dynamics and spin diffusion, so
we assume quasi-neutrality in the channel at all times ($\delta
n_{+}(t)$$+$$\delta n_{-}(t)$$=$$0$). In the linear regime under
consideration (when $\Delta \xi$$<$$k_{B}T$) the average
electrochemical potential $\xi$$=$$(\xi_{+}+\xi_{-})/2$ is equal to
$-e\phi$. At every moment of time $\xi$ fulfills the Laplace
equation with boundary conditions given by currents at the
interfaces. In the time-dependent case these include also
displacement currents connected with charging of the barrier
capacitance $C_{B}$. A Schottky barrier is a dipole layer, and its
capacitance can have a strong effect on dynamics of currents on time
scales of interest here. With the displacement current taken into
account, the boundary condition for spin current is:
\begin{equation}
j_{s} = \frac{G_{s}}{e}(\mu^{m}(t) - \xi_{s}(t)) + \frac{c_{B}}{2e}
\frac{\partial}{\partial t}(\mu^{m}(t) - \xi(t) ) \,\, ,
\label{eq:js_displacement}
\end{equation}
where $c_{B}$ is the barrier capacitance per unit area. The second term in the above equation
represents the carriers which flow towards the barrier, but do not
tunnel through it. Instead, they stay in the semiconductor close to
the barrier, making the depletion region slightly thinner or wider.
The charge involved in this process is negligible compared to the
charge already swept out from the semiconductor, so we can keep
$c_{B}$ constant. For small spin splitting (so that the
conductivities $\sigma_{+}$$\simeq$$\sigma_{-}$) the same amount of
carriers of each spin is going to be brought from the channel into
the barrier, and the displacement current is the same for each spin
in Eq.~(\ref{eq:js_displacement}). For layer-averaged $\xi$ we get
then
\begin{equation}
\frac{ \partial^{2} \xi}{\partial x^{2}} =
-\frac{\alpha_{i}}{2L^{2}}(\mu^{m}_{i}-\xi)
+\frac{\beta_{i}(t)}{4L^2} \Delta \xi - \frac{c_{B}}{\sigma
h}\frac{\partial}{\partial t} (\mu^{m}_{i}-\xi)  \,\, ,
\label{eq:xi_displacement}
\end{equation}
with the right hand side of Eq.~(\ref{eq:xi_displacement})
non-zero only under the contacts.  The magnetization dynamics of
$i^{th}$ contact  translates into time-dependence of $\beta_{i}$,
driving the spin diffusion in Eq.~(\ref{eq:Dxi}) and electric
potential in the channel in Eq.~(\ref{eq:xi_displacement}). From
$\xi_{s}$ we calculate the current $I_{R}(t)$ charging the capacitor
$C$. The electrochemical potential of the R terminal
$\mu_{R}$$=$$-eV_{R}$ changes according to  $dV_{R}/dt$$=$$I_{R}/C$.
Examples of calculations for two possible modes of operation
(sensing the L dynamics and reading out the L/M alignment) are shown
in Fig.~\ref{fig:capacitor_current}.

\begin{figure}[t]
\begin{center}
\includegraphics[width=8cm]{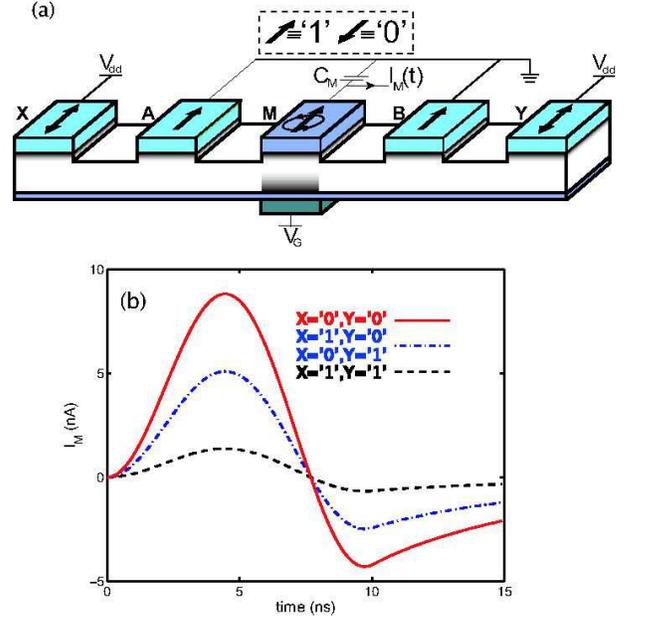}
\caption{(a) A five-terminal magneto-logic gate (MLG). The logic
inputs `0' and `1' are encoded by magnetization direction of the
A,B,X,Y terminals (see text for details). As shown here, the gate is
set to work as a NAND operation between X and Y (A and B are fixed
to `1' values). In the read-out phase the magnetization of the
middle (M) terminal is rotated by $2\pi$, or the back-gate voltage
$V_{G}$ is increased. (b) The $I_{M}(t)$ transient current triggered
by the rotation of the M magnetization. The small signal for X=`1', Y=`1' corresponds to logical `0' output. Adapted from Ref.~38.} \label{fig:MLG}
\end{center}
\end{figure}

\subsection{Reprogrammable Magneto-Logic Gate} \label{sec:MLG}
The same physical principle of operation can be harnessed to achieve
a higher level functionality. In Fig.~\ref{fig:MLG}a we present
a scheme of a five-terminal system,\cite{Dery_Nature07} in which the
electric sensing of spin accumulation is used to perform a logic
operation, i.e., two bits of input are converted into a binary
output signal. This is a reprogrammable magneto-logic gate (MLG).
Spintronics logic gates have been proposed in purely metallic
systems,\cite{Cowburn_Science00,Hanbicki_APL01,Richter_APL02,Ney_Nature03,Allwood_Science05,Imre_Science06}
but ours is the first proposal which employs semiconductors as
active elements of the system.

The system presented in Fig.~\ref{fig:MLG}a works in the following
way. The charge currents are flowing between two pairs of terminals
(X and A, Y and B), between which the bias $V_{dd}$ is applied.
Depending on the alignment of these pairs of magnets, different
patterns of spin accumulation are created in the channel: if both
X/A and Y/B are AP the spin accumulation underneath M is large; if
only one pair of contacts is AP the spin accumulation is
approximately two times smaller, and if both pairs are P there is a
very small $\Delta \xi$ beneath the M terminal. The M contact is
used to directly express the differences in the average spin
accumulation beneath it.

The logical inputs are encoded by magnetization directions of A, B,
X, and Y terminals. We will concentrate on the case in which A and B
magnetizations are preset, defining the logic function of the gate.
This reprogrammability is an important feature of
magnetization-based logic. X and Y are then the logic inputs, and
the output is generated when the M magnet is rotated by $2\pi$,
triggering a transient $I_{M}(t)$ current of amplitude proportional
to the spin accumulation in the middle of the channel. Let us focus
on the example of the NAND gate, as any other logic function can be
realized by using a finite number of such gates. For the NAND
operation, A and B magnets are set parallel to each other in the
direction defining the logical `1'. The amplitude of the $I_{M}(t)$
oscillation is two times larger for $X$$=$`0' and $Y$$=$`0' compared
to the case when one of them is `0' and the other is `1', and for
`11' case the current is negligible. This is shown in
Fig.~\ref{fig:MLG}b. The transient current can be captured by an
external electronic circuit, and then used to control a suitable
write operation applied to the magnetic contact of another
gate.\cite{Dery_Nature07}

Instead of using magnetic field pulses to drive the $2\pi$ rotation
of the M magnet, one can employ the idea of the spin switch outlined
in Sec.~\ref{sec:extraction}, in which the M magnetization is
pinned, and the profile of the conduction band beneath M is changed
by applying a voltage pulse $V_{G}(t)$ to the back gate shown in
Fig.~\ref{fig:MLG}a. $V_{G}$ is chosen to deplete the electrons
from the lower part of the channel, biasing them vertically towards the Schottky barrier. Thus, the effective spin
selectivity of the M junction is switched as discussed in
Sec.~\ref{sec:extraction}. This is qualitatively the same as
reversing the M magnetization, and the transient current $I_{M}$
should be generated. However, its quantitative calculation is much
more involved than in the case of magnetization rotation, and it
will be addressed in future work.

\section{Summary}
We have reviewed the physics of spin injection and extraction
through inhomogeneously doped Schottky barriers. We have paid
special attention to the states localized close to the junction as a
consequence of heavy doping near the interface, and we have shown
that they play a crucial role in the recent spin extraction experiments.
A junction held at low bias in the presence of spin accumulation in
the semiconductor has been also analyzed, and how
its spin-dependent properties could be used to electrically sense
the spin accumulation in a semiconductor has been described. The multi-terminal systems
(a three-terminal transistor and a five-terminal logic gate) which
we have described rely only on spin-selectivity of the junctions and
the presence of the spin accumulation. As such, they are ideal
candidates for spintronics devices working at room temperature
and, possibly, in silicon.

\section*{Acknowledgments}
This work was supported by the NSF under Grant No.~DMR-0325599. We
are grateful to P. Crowell, M. Johnson,
and B. Jonker for helpful discussions.




\end{document}